\documentclass[aps,prl,showpacs,twocolumn,superscriptaddress]{revtex4-1}
\usepackage{amsfonts}
\usepackage{graphicx}
\usepackage{dcolumn}
\usepackage{bm}

\newcommand{\nc}{\newcommand}
\nc{\be}{\begin{equation}}
\nc{\ee}{\end{equation}}
\nc{\bea}{\begin{eqnarray}}
\nc{\eea}{\end{eqnarray}}
\nc{\bean}{\begin{eqnarray*}}
\nc{\eean}{\end{eqnarray*}}
\nc{\mb}{\mbox}
\nc{\rnc}{\renewcommand}
\nc{\vk}{\mb{\bf k}}
\nc{\vp}{\mb{\bf p}}
\nc{\vn}{\mb{\bf n}}
\nc{\vq}{\mb{\bf q}}
\nc{\rr}{\mb{\bf r}}
\nc{\vz}{\hat {\mb{\bf z}}}
\nc{\vj}{\mb{\boldmath$j$}}
\nc{\vg}{\mb{\boldmath$g$}}
\nc{\x}{\mb{\boldmath$x$}}
\nc{\A}{\mb{\boldmath$A$}}
\nc{\va}{\mb{\boldmath$a$}}
\nc{\vs}{\mb{\boldmath$\sigma$}}
\nc{\vpi}{\mb{\boldmath$\pi$}}
\nc{\nab}{\nabla}
\nc{\X}{\sf x}

\begin{document}
\title{ Collective modes in singlet superconductors }
\author{Yafis Barlas}
\affiliation{Department of Physics and Astronomy, University of California,
Riverside, CA 92521}
\author{C. M. Varma}
\affiliation{Department of Physics and Astronomy, University of California,
Riverside, CA 92521}

\begin{abstract}
{A brief summary of collective mode excitations that can 
exist in singlet superconductors with irreducible representation $L$ is given. Such excitations may be classified as the coupled excitations of the charge density $\rho$ and the phase $\phi $ of the order parameter, or of the amplitude $\Delta$ of order parameter.
Each of these classes may be further characterized in the long wavelength limit  by the irreducible representation $\ell$ of the excitation, which may or may not be the same as the ground state $L$.}
\end{abstract} 
\maketitle

In describing a state of superconducting order and its excitations, it is useful to express the Hamiltonian in the form $\mathcal{H} 
= \mathcal{H}_{BCS} + 
\mathcal{H}_{int}$ with
\begin{equation}
\mathcal{H}_{BCS} = \sum_{{\bf k}} \psi^{\dagger}_{{\bf k}} (\epsilon_{k} \hat{\tau}_{3} +
\Delta_{{\bf k}} \hat{\tau}_{1} ) \psi_{{\bf k}}, 
\end{equation}
where we take $\Delta_{{\bf k}}$ real, $\hat{\tau}_{i} $ are the Pauli
matrices $i=1,2$ and $3$ and $\psi^{\dagger}_{{\bf k}} = (c^{\dagger}_{{\bf k} \uparrow}, 
c_{-{\bf k} \downarrow})$ is the Gorkov spinor.  We will assume that the superconducting order 
parameter $\Delta_{{\bf k}}$ is in a given irreducible representation $L$ (for s-wave superconductors $L = 0$). The Hamiltonian 
describing the residual interaction can be expressed as 
\begin{eqnarray}
\nonumber
\mathcal{H}_{int} &=& \frac{1}{2} \sum_{{\bf k},{\bf k}',{\bf q}} v_{{\bf q}} ({\bf k, k}')
(\psi^{\dagger}_{{\bf k}+ {\bf q}} \hat{\tau}_{3} \psi_{{\bf k}})
(\psi^{\dagger}_{{\bf k}'- {\bf q}} \hat{\tau}_{3} \psi_{{\bf k}'}) \\
&-& \sum_{{\bf k}} \Delta_{{\bf k}} \psi^{\dagger}_{{\bf k}} \hat{\tau}_{1}
\psi_{{\bf k}},
\end{eqnarray}
where $v_{{\bf q}}({\bf k, k}')$ is the renormalized density-density interaction that includes 
the effect of phonons, spin fluctuations or other excitations of relevance to superconductivity besides the Coulomb interactions.\newline 
\indent
In considering collective fluctuations, which are effects beyond the BCS approximation,  $\mathcal{H}_{int}$ may be expressed in  a separable form for ${\bf k,k}'$ near the Fermi-surface. 
\begin{equation}
v({\bf q})({\bf k, k}') = \sum_{\ell} V_{l} F_{\ell}({\bf q})  P_{\ell} (\theta) P_{\ell} (\theta').
\end{equation}
 For the $\ell=0$ channel, $F_{\ell}({\bf q})$ must be taken to be the Coulomb interaction in order to correctly treat the longitudinal charge density excitation of the superconductor. For other $\ell$, $F_{\ell}({\bf q})$ is a short-range interaction. \newline
\indent
In conventional s-wave superconductors there are three types of 
long-wavelength fluctuations: the charge-density fluctuations $\rho$, the 
fluctuation of the phase $\phi$ and the fluctuation in the amplitude
of the superconducting order parameter $\Delta$.  In the Gorkov spinor basis 
charge density fluctuations appear in the $\hat{\tau}_{3}$ scattering channel, the 
phase fluctuations in the $\hat{\tau}_{2}$ scattering channel, and the amplitude fluctuations appear 
in the $\hat{\tau}_{1}$ scattering channel. The phase fluctuations are coupled to the density fluctuations for a
 gauge invariant theory, i.e. a theory consistent with the continuity equation. For any $L$, there then exists a coupled mode in the $\ell=0$ channel which in the long wavelength ($ q \to 0$) limit is close (with order $|\Delta/E_f|^2$) to the plasma frequency of the normal metallic state~\cite{Anderson}. For convenience, we list the nature and the eigenvectors of the excitations in terms of bare fermion operators, in the $ q \to 0$ limit is given in the table below:
\begin{displaymath}
\begin{array}{  c | c | c  }
 & {\rm channel} & {\rm eigenvectors} \\
\hline
{\rm density} & \hat{\tau}_{3} &  e \sum_{\sigma} c^{\dagger}_{k,\sigma} c_{k,\sigma}\\
{\rm phase} & \hat{\tau}_{2} & c^{\dagger}_{k,\sigma} c^{\dagger}_{-k,-\sigma} - c_{k,\sigma}c_{-k,-\sigma}  \\
{\rm amplitude} & \hat{\tau}_{1} & c^{\dagger}_{k,\sigma} c^{\dagger}_{-k,-\sigma} + c_{k,\sigma}c_{-k,-\sigma} \\
\end{array}
\end{displaymath}
We will denote by  $\ell$ to the relative angular momentum associated with the 
excitations in the long wavelength limit. \newline
\indent
As already mentioned, the $\hat{\tau}_{3}$ and the $\hat{\tau}_{2}$ 
channels are coupled and due to the long-range part of the Coulomb 
interaction in the $\ell=0$ interaction channel, and are therefore 
pushed up to the plasmon energy~\cite{Anderson}. (This is also the 
mechanism whereby W and Z bosons were proposed for a unified model 
of weak and electromagnetic interactions.\cite{Weinberg} \newline
\indent
For s-wave ($L=0$) superconductors Bardasis and Schriffer~\cite{BS} 
argued that because  $\ell \ne 0$ interactions are of finite range, there may be bound states for 
$\tau_3$ excitations in such angular momentum channels below $2 \Delta$. Such modes can in principle directly couple to external probes in the infra-red (odd $\ell$) or Raman (even $\ell$) experiments. These modes have also been discussed explicitly by Martin~\cite{Martin} and by 
Tutto and Zawadawski~\cite{TZ}. Experiments have not given evidence for the existence of such modes.  \newline
\indent
Modes in the $\hat{\tau}_{3}$ channel are also of-course possible for $L \ne 0$ ground states. Again the $\ell =0$ mode will be the plasmon. The $\ell \ne 0$ modes, were they to exist below $2\Delta$ would however overlay a continuum of incoherent particle-hole excitations generally characteristic of $L \ne 0$ pairing. The situation may be different for extended s-pairing due to complicated band-structures, which may have a gap to the incoherent excitations. Such a possibility has been recently discussed
by Scalapino and Devereaux~\cite{SD}.\newline
\indent
It is instructive to note that the different $\ell$ excitations in the $\hat{\tau}_{3}$ channel correspond in the normal state, irrespective of the superconductivity $L$-state, to what may be termed the Landau-Pomeranchuk modes. In other words, 
in the long-wavelength ($q \to 0$) limit, they correspond to distortion of the
Fermi surface in different angular momentum channels. These are in general broad in energy due to mixing with incoherent excitations. In the superconducting state 
due to the presence of the gap, they may have bound state parts.  Klein and 
Dierker~\cite{KD} considered the effect of the penetration depth of light which can lead to 
mixing of angular momentum channels. But just as the $\ell =0$ modes move to plasmons in the bulk, the surface boundary conditions can only transform them to a surface plasmon excitations just as in the normal state.\newline
\indent
We now discuss the $\hat{\tau}_{1}$ or the amplitude modes in the 
superconductors, introduced by Littlewood and Varma~\cite{LV}. They were 
derived \cite{LV} as a pole in the Bethe-Salpeter 
equation in the $\hat{\tau}_{1}$ or amplitude channel. 
They arise only \cite{Varma} because the superconducting state is particle-hole 
symmetric at low energies to an accuracy of $O(\Delta/E_f)$. These modes have the same physics as the Higgs mode in particle physics \cite{Higgs}. They could have been derived long ago from the Ginzburg-Landau form of free-energy if the dynamical equations were considered with second order in time or particle-hole symmetric and not first order in time as was usually the case. For $L=0$ superconductors and in the $\ell = 0$ channel, 
 this mode has the energy $2 \Delta$ for $q \to 0$. This is because this channel is orthogonal to the density channel and therefore the Coulomb interactions do not affect it. \newline
\indent 
There is an important distinction between the $\hat{\tau}_{1}$  mode and the coupled $\hat{\tau}_{3}-\hat{\tau}_{2}$ modes which allows easy experimental distinction between them. Being chargeless and spin-less, no external probe can couple to them ${\it directly}$. However such modes couple linearly to excitations which shake the superconducting ground state. If the latter couples directly to the external probes, the $\hat{\tau}_{1}$ modes must then steal the spectral weight from them~\cite{LV}. The relative magnitude of this effect diminishes rapidly with the difference of the energy of the directly coupled modes from $2\Delta$.  This phenomena was shown to be consistent with observations in $NbSe_{2}$. In $NbSe_{2}$ the phase transition 
to superconductivity at 7.2 K is preceded by a charge-density-wave (CDW) transition at 45 K.
The CDW phonon modulates the CDW amplitude which 
couples to superconducting order parameter by changing the number of 
electron available for superconducting order. Furthermore, due to 
the close proximity of the phonon pole (which appears at $\approx 5 
\Delta$) this amplitude mode can be observed in the renormalized phonon 
propagator by Raman spectroscopy. \newline
\indent
In experiments, one can distinguish the $\hat{\tau}_{1}$ and the coupled $\hat{\tau}_{3}-\hat{\tau}_{2}$ modes for any $L$ and $\ell$. The former can be seen only through a corresponding diminution in the intensity of some other modes, while the latter have no such requirement.
For s-wave superconductors $L=0$, one can also derive $\hat{\tau}_{1}$ excitations in relative angular 
momentum channels $l \neq 0$. This can be done by looking for a pole 
in the Bethe-Salpeter equation in the angular 
momentum channel. We do not know of any experiments which have seen these.\newline
\indent
In our current work~\cite{BV} we have extended the concept of this $\hat{\tau}_{1}$
amplitude mode to ground states with reduced point group symmetry, 
such as d-wave superconductors $ L =2$. We analyze the excitations of the order 
parameter in different irreducible representations $\Gamma_{i}$ of the point group 
symmetry of the lattice by expressing 
\begin{equation} 
\Delta_{{\bf k}} = \sum_{\Gamma_{i}} \Delta_{\Gamma} (\Gamma_{i}).
\end{equation}
 These amplitude modes are orthogonal 
to the $\hat{\tau}_{3}-\hat{\tau}_{2}$ or Bardasis-Schrieffer type modes. In our current work~\cite{BV} we have explored 
consequences of these excitations. In order to put our predictions in perspective we have summarized the known collective 
modes in superconductors in the form of a table (see the Appendix below).

\begin{widetext}
\section{Appendix : Collective modes in singlet superconductors}
\begin{displaymath}
\begin{array}{  c | c | c | c }
{\rm modes} & {\rm pairing} & {\rm relative} & {\rm energy} \\
  & {\rm momentum} & {\rm momentum} &   \\
\hline
{\rm phase}, \phi &  L = 0 &    & \omega_{p}  \\
\hline
{\rm density} & L = 0 &  l=0 & \omega_{p}  \\
\rho  & L = 0 &  l \neq 0 & < 2\Delta   \\
    &  L \neq 0 &  l=0 & \omega_{p} \\
    & L \neq 0 &  l \neq 0  & < 2\Delta   \\
\hline
{\rm amplitude} & L = 0 &  l=0  & 2 \Delta  \\
\Delta  & L = 0 &  l \neq 0 &     < 2\Delta  \\
    &  L \neq 0 &  l=0 & 2 \Delta  \\
    & L \neq 0 &  l \neq 0 &   < 2\Delta   \\
\end{array}
\end{displaymath}

\end{widetext}

\end{document}